\documentclass{jpsj2}
\title{Charge-density-wave instability in the Holstein model with quartic anharmonic phonons}

\author{\textsc{Alain~Mo\"ise~Dikand\'e}\thanks{E-mail: adikande@ictp.it}}
\inst{Laboratory of Research on Advanced Materials and Nonlinear Sciences(LaRAMaNS), Department of Physics, Faculty of Science, University of Buea PO Box 63 Buea, Cameroon}
\abst{The molecular-crystal model, that describes a one-dimensional electron gas interacting with quartic anharmonic lattice vibrations, offers great potentials in the mapping of a relatively wide range of low-dimensional fermion systems coupled to optical phonons onto quantum liquids with retarded interactions. Following a non-perturbative approach involving non-Gaussian partial functional integrations of lattice degrees of freedom, the exact expression of the phonon-mediated two-electron action for this model is derived. With the help of Hubbard-Stratonovich transformation the charge-density-wave instability is examined in the sequel, with particular emphasis on the effect of the quartic anharmonic phonons on the charge-density-wave transition temperature.}
\kword{Holstein model, anharmonic phonons, non-Gaussian functional integrals, Parabolic Cylinder functions, Hubbard-Stratonovich transformation, charge-density-wave correlations}
\begin{document}
\maketitle
\section{Introduction} 
Standard electron-phonon models~\cite{frohlich,su,holstein} have served as a rich workshop for studies of phase transitions in one-dimensional($1D$) systems with coexisting structural instabilities and electronic correlations. However, while these standard models provide relevant qualitative insights onto physical mechanisms of the phase transitions, potential materials reported to date display rather diverse lattice dynamics as well as a great variety of local structures hence bringing about the fundamental question of their universality~\cite{bish2,bish4,taka}. On another hand, failures of traditional formalisms such as the Hartree-Fock and weak-coupling expansions of the classic electron-phonon interactions, paved the way towards a many-body picture~\cite{luther,cross} of structural instabilities in low-dimensional electronic as well as magnetic chains. The key asset of many-body approaches resides in the possibility to map the $1D$ electron-phonon problem onto a $1D$ interacting electron gas~\cite{schulz,schulza,schulzb,schulz1,voit} with retarded electron-electron interactions, so that well-developed quantum-field theoretical concepts for low-dimensional electronic liquids can also apply in these specific contexts\cite{luther,cross,dika3}. In particular, this mapping revealed the inconsistency of the quasiparticle picture inherent to one-electron perturbative approaches, and permitted a better understanding of several experimental observations one must fascinating being the non-universal character of critical exponents of correlation functions~\cite{schulz,schulza,schulzb,schulz1,voit,schulz2}. \\
The present study aims at investigating the charge-density-wave(CDW) instability for the molecular-crystal model with quartic anharmonic phonons, with emphasis on the contribution of phonon anharmonicity to the transition temperature. A key step in this respect will be an exact formulation of the effective two-electron action for this model which is valid at any arbitrary regime(i.e. weak or strong) of crystal anharmonicity. Instructively, the issue of the interplay of electrons and anharmonic phonons has been considered recently~\cite{bish2,bish4,taka,jarell1,jarell2,nasu1,nasu2,grzyb} in attempts to explain unsual local structure effects on thermal and spectral properties of certain materials~\cite{taka,nasu1,nasu2}. In particular, refs.~\cite{jarell1,jarell2} have addressed the specific problem of phonon anharmonicity and its effects on the superconducting and CDW transition temperatures. Thus, following a self-consistent Gaussian decoupling of the quartic anharmonic phonon, it has been established that the electron-anharmonic phonon model could be reduced to an effective electron-harmonic phonon one without the particle-hole symmetry, but with a persistent finite transition temperature at half filling of the electronic band. On the basis of this self-consistent Gaussian treatment, the authors found that the phonon anharmonicity had no qualitative effect on electronic instabilities in anharmonic crystals except breaking the particle-hole symmetry. \\
Here we wish to extend results of refs.~\cite{jarell1,jarell2} for the CDW instability, to materials with relatively strong quartic anharmonicities. In this purpose, we shall establish that by proper account of phonons via non-Gaussian partial functional integrations~\cite{witch,tus} of phonon degrees of freedom, one can obtain the exact retarded two-electron action for the anharmonic molecular-crystal Hamiltonian. By use of appropriate Hubbard-Stratonovich fields, an effective field theory for the CDW instability will be constructed and the transition temperature derived. 
\section{The anharmonic-phonon-induced two-electron action}
\begin{figure}[tb]
\begin{center}
\includegraphics{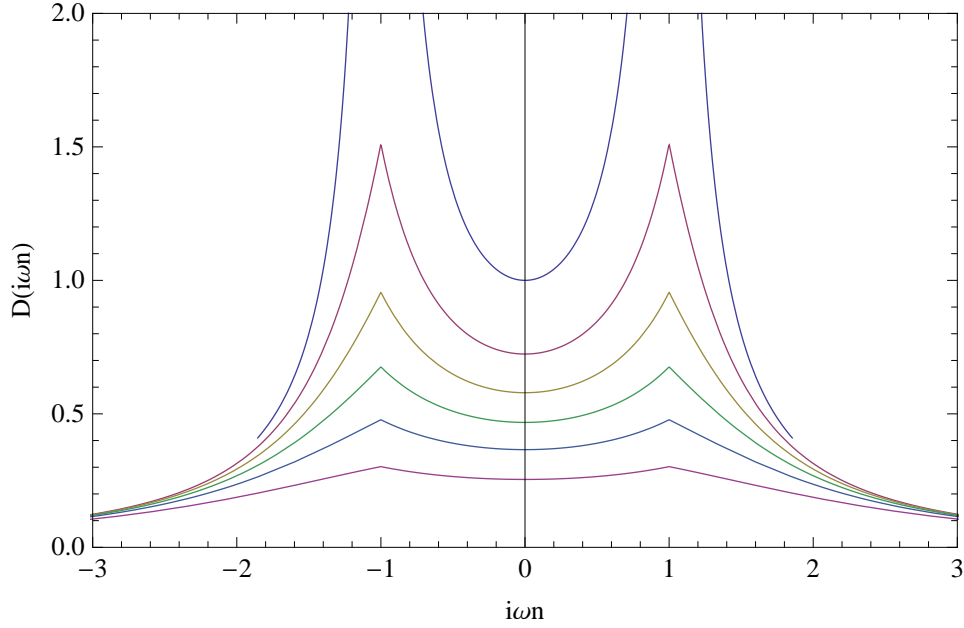}
\end{center}
\caption{The reduced one-phonon propagator versus $i\omega_m/\omega_0$, for $\lambda_{eff}=0$, 0.2, 0.5, 1, 2, 5. The diverging curve corresponding to $\lambda_{eff}=0$, coincides exactly with the reduced free-phonon propagator $\mathcal{D}_0(\omega_m)/\mathcal{D}_0(0)$.}
\label{f1}
\end{figure}

\begin{figure}[tb]
\begin{center}
\includegraphics{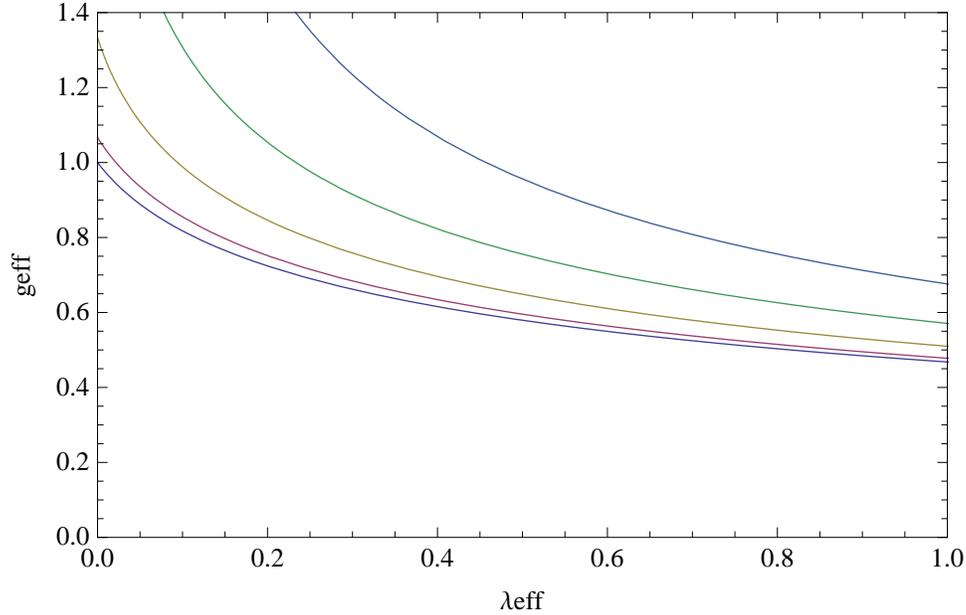}
\end{center}
\caption{Variation of the reduced effective electron-electron interaction $g_{\lambda}(0,T)/g_o(0)$ as a function of the effective anharmonicity coefficient $\lambda_{eff}$, for $i\omega_m=0$, $0.25\,\omega_0$, $0.5\,\omega_0$, $0.75\,\omega_0$ and $\omega_0$(from bottom to top curves).}
\label{f2}
\end{figure}

The standard Holstein model~\cite{holstein} describes a $1D$ electron gas coupled to a molecular lattice with quasi-optical harmonic vibrations. For crystal lattices dominated by quartic anharmonic vibrations, this model can be extended leading to an anharmonic Holstein model which, at half-filling of the electronic band, reads:
\begin{eqnarray}
H&=& \sum_{j=1}^{N}{\left[\frac{P_j^2}{2M} + \frac{M\omega_o^2}{2}\,r_j^2 + \lambda\,r_j^4\right]} \nonumber \\
 &-& t\,\sum_{j=1,\sigma}^{N}{{c_{j,\sigma}}^{\dagger}c_{j',\sigma}} + \alpha\sum_{j=1,\sigma}^{N}{r_j \,{c_{j,\sigma}}^{\dagger}c_{j,\sigma}}, \label{a1}
\end{eqnarray}
where $r_j$ is the displacement of the $j^{th}$ molecule of mass $M$ and characteristic frequency $\omega_o$ relative to its equilibrium, $P_j$ is the momentum conjugate to $r_j$ and $\lambda$ is the coefficient of anharmonicity. $c_{j,\sigma}^{\dagger}$($c_{j,\sigma}$) creates(annihilates) an electron state of spin $\sigma$ at site $j$ and $t$ is the uniform transfer integral between nearest-neighbour sites $j$ and $j'=j\pm 1$. Lastly, $\alpha$ is the electron-phonon interaction coefficient. \\ 
The Holstein model has a long and rich history in the literature of polarons in $1D$ and $2D$ materials, high-temperature cuprate superconductors and colossal magnetoresistance manganites~\cite{salje,taka2,taka3,taka4,taka5} being two classes of materials that attracted much recent attention because of a sizable signature of polaronic effects in their heat and charge transport parameters. As for electronic instabilities, this model has been shown to possess a rich phase diagram covering CDW and superconductivity as observed in $Ba_{1-x}K_xBiO_3$, a material that exhibits a CDW-driven metal-insulator transition and which upon doping becomes superconducting. From the standpoint of fundamental physics, the Holstein model is particularly interesting as it represents one among the few rare models in which a net departure from Fermi liquid behaviour occurs. Indeed, there the electronic spectral density looses the quasiparticle peak and is swamped by an incoherent background with increasing electron-phonon coupling~\cite{eng,marsiglio}. In general these properties can be unveiled either analytically as in the case of the $1D$ Holstein model with a linearized electronic dispersion which is exactly solvable, or via quantum Monte-Carlo and diagrammatic approaches based on Migdal-Eliashberg theory as well established~\cite{marsiglio,scatt,marsiglio1,noack1,noack2,noack3,white} for the $2D$ Holstein model. At half filling of the electronic band and strong electron-phonon coupling, the model gives rise to an effective Hubbard model so that CDW and on-site s-wave superconductivity coexist. Away from half filling or with decreasing electron-phonon coupling but high phonon frequencies, the superconducting phase dominates. In recent years there has been much interest to the Holstein-Hubbard model~\cite{grilli,huang,koller,san1,san2} for which the competition between phonon-mediated and direct electronic correlations, permits the account of new groundstates in the phase diagram of the bare Holstein model including spin-density-wave and d-wave superconducting phases. Remarkably, the functional formalism involving partial integrations of phonons has demonstrated great effectiveness in this last context by furnishing the retarded attractive two-electron Hamiltonian competing with Hubbard's Coulomb repulsion~\cite{schulza,schulzb,voit}. \\
 In this work we address the issue of a possible exact formulation of the phonon-mediated two-electron action for the Holstein model with quartic anharmonic phonons~(\ref{a1}), given its importance that extends far beyond the specific context of CDW instability considered in the present context if looked out from the more global standpoint of the effective-field theory for electronic instabilities within the framework of the Holstein-Hubbard model~\cite{schulz,schulza,schulzb}. Partially integrating out lattice degrees of freedom in the total Hamiltonian~(\ref{a1}), we arrive at the following partition function in the Matsubara space and field-theory representation~\cite{dumoulin}:
\begin{equation}
Z= Z_{l}^0\,\int{\prod_{k,\sigma,\tau} D\psi_{k,\sigma}(\tau)\,D\psi_{k,\sigma}^{\star}(\tau)\,\,e^{\lbrack S_{e}^0(\psi,\psi^{\star}) + S_I(\psi,\psi^{\star})\rbrack}}, \label{a2}
\end{equation}
where 
\begin{equation}
Z_{l}^0=\int{\prod_{q,\tau} Dr(q,\tau)\, e^{S_{l}^0(r)}}  \label{a3} 
\end{equation}
is the bare lattice partition function corresponding to the anharmonic action: 
\begin{eqnarray}
S_{l}^0&=& \frac{1}{2}\sum_q{\int_0^{\beta}{d\tau\, \mathcal{D}_0^{-1}(\tau)\,r(q,\tau) r(-q,\tau)}} \nonumber \\ 
&-& \frac{\lambda}{L} \,\sum_{q_1,q_2,q_3,q_4}{\int_0^{\beta}{d\tau\, r(q_1,\tau) r(q_2,\tau) r(q_3,\tau) r(q_4,\tau)}}, \nonumber \\ 
\label{a4}
\end{eqnarray}
$L=Na$ being the total length of the system. In formula~(\ref{a4}), $r$ is the bosonic degree of freedom in the field configuration for the classical displacement $r$ i.e. 
\begin{equation}
r(x,\tau)=\frac{1}{\sqrt{L}}\sum_q{r(q,\tau)\exp(iqx)}, \hskip 0.2truecm x\equiv j\,a, \label{a5}
\end{equation}
while $\mathcal{D}_o(\tau)$, the free-phonon propagator, is derived from the general expression of the retarded one-phonon propagator i.e.  
\begin{equation}
\mathcal{D}_{\lambda}(\tau)= -<T_{\tau}\,r(q,\tau)\,r(-q, 0)>_{\lambda},
\end{equation}
in the Gaussian limit $\lambda\rightarrow 0$. In the Matsubara space $\mathcal{D}_o(\tau)$ becomes: 
\begin{equation}
\mathcal{D}_0(\omega_m)= -\frac{M^{-1}}{\omega_m^2 + \omega_o^2}, \label{a6}
\end{equation}
with $\omega_m= 2\pi m\beta^{-1}$ the Matsubara frequency. Lastly, 
\begin{equation}
S_{e}^0= \sum_{k,\sigma}{\int_0^{\beta}{d\tau\,\psi_{k,\sigma}^{\star}(\tau)G_0^{-1}(k,\tau)\psi_{k,\sigma}(\tau)}}   \label{a7}
\end{equation}
is the free-electron action where
\begin{equation}
G_0(k,\tau)=-\partial/\partial \tau - \upsilon_F k \label{a7a}
\end{equation}
is the free-electron propagator for an electronic dispersion linearized around the Fermi wavector $k_F$($\upsilon_F$ being the Fermi velocity), and $\psi$ is the usual Grassmann field. 
\\ The quantity in~(\ref{a2}) of primary interest to us is the phonon-mediated effective two-electron action, i.e.: 
\begin{equation}
S_I(\psi,\psi^{\star})= \ln\left[< e^{S_{e-l}(\psi,\psi^{\star};\,r)}>_{\lambda} \right], \label{a8}
\end{equation}           
in which
\begin{equation}
S_{e-l}(\psi,\psi^{\star};\,r)=
\frac{\alpha}{\sqrt{L}} \sum_{k,q,\sigma}{\int_0^{\beta}{d\tau\, \psi_{k+q,\sigma}^{\star}(\tau)\psi_{k,\sigma}(\tau)r(q,\tau)}}  \label{a9}
\end{equation}
is the electron-lattice interaction action with averages over the anharmonic lattice action. Since the lattice action is anharmonic in the phonon fields, a Gaussian treatment of averages over phonons requires that the anharmonic coefficient is very weak compared to the harmonic stifness $M\omega_0^2$~\cite{schneider}. However, if the two coupling coefficients are strongs the above perturbation consideration will inevitably result in a crude approximation of the problem. Actually, only by exact integration of phonon degrees of freedom can one gain full understanding of both qualitative and quantitative contributions of the phonon anharmonicity to the retardation process in the resulting effective two-electron action. In this last purpose, assume for now that the quadratic mean over boson fields in~(\ref{a8})-(\ref{a9}) can be calculated exactly and leads to an effective interaction action which we formally represent as:    
\begin{equation}
S_I= -\frac{1}{2L}\,\sum_{k,k',q,\sigma,\sigma'}{\int_0^{\beta}{d\tau' \int_0^{\beta}{d\tau g_{\lambda}(\tau-\tau')\,\psi_{k+q,\sigma}^{\star}(\tau)\psi_{k'-q,\sigma'}^{\star}(\tau')\psi_{k',\sigma'}(\tau')\psi_{k,\sigma}(\tau)}}}. \label{a10}
\end{equation}
In the last formula the phonon-mediated electron-electron interaction matrix has been defined as:
\begin{equation}
g_{\lambda}(\tau-\tau')= \alpha^2 \mathcal{D}_{\lambda}(\tau-\tau'), \label{a11}
\end{equation}
where the quantity $\mathcal{D}_{\lambda}(\tau-\tau')$ is the exact one-phonon propagator in the presence of quartic anharmonicity. Following conventional notations, we define the Matsubara Fourier transform of $\mathcal{D}_{\lambda}(\tau-\tau')$ as:
\begin{equation}
\mathcal{D}_{\lambda}(\tau-\tau')= \beta^{-1}\sum_{m=-\infty}^{\infty}{\mathcal{D}_ {\lambda}(\omega_m)\,e^{-i\omega_m\,(\tau-\tau')}}, \label{a12}
\end{equation}
and with the help of a standard non-Gaussian integral~\cite{witch,tus}, the anharmonic one-phonon propagator $\mathcal{D}_{\lambda}(\omega_m)$ is exactly found:
\begin{eqnarray}
 \mathcal{D}_{\lambda}(\omega_m)&=& -\frac{1}{\sqrt{2\tilde{\lambda}}}\,\frac{D_{-3/2}\lbrack \frac{-1}{\sqrt{2\tilde{\lambda}}}\mathcal{D}_0^{-1}(\omega_m)\rbrack}{D_{-1/2}\lbrack \frac{-1}{\sqrt{2\tilde{\lambda}}}\mathcal{D}_0^{-1}(\omega_m)\rbrack}, \nonumber \\
 \tilde{\lambda}&=& \lambda\, T/L, \label{a13}
\end{eqnarray}
with $D_p(y)$ the Parabolic Cylinder function of order $p$ and argument $y$~\cite{hand1}. With help of the explicit formula of $\mathcal{D}_{\lambda}(\omega_m)$ just derived, we transform the two-electron action $S_I$ in the Matsubara Fourier space obtaining:
\begin{equation}
S_I= -\frac{T}{2L}\,\sum_{\tilde{k},\tilde{k}',\tilde{q},\sigma,\sigma'}{g_{\lambda}(\omega_m, T)\,\psi_{\tilde{k}+\tilde{q},\sigma}^{\star}\psi_{\tilde{k}'-\tilde{q},\sigma'}^{\star}\psi_{\tilde{k}',\sigma'}\psi_{\tilde{k},\sigma}} \label{a14}
\end{equation}
where $\tilde{k}\equiv (k,\omega_n)$, $\tilde{q}\equiv (q,\omega_m)$ and the phonon-mediated electron-electron interaction matrix now expressed in Matsubara Fourier space reads:
\begin{equation}
g_{\lambda}(\omega_m, T)= -\alpha^2 \, \frac{1}{\sqrt{2\tilde{\lambda}}}\,\frac{D_{-3/2}\lbrack \frac{-1}{\sqrt{2\tilde{\lambda}}}\mathcal{D}_0^{-1}(\omega_m)\rbrack}{D_{-1/2}\lbrack \frac{-1}{\sqrt{2\tilde{\lambda}}}\mathcal{D}_0^{-1}(\omega_m)\rbrack}. \label{a15} 
\end{equation}
To see what formula~(\ref{a15}) becomes in the harmonic-phonon regime, remark that when $\lambda\rightarrow 0$ the argument of the Parabolic Cylinder function becomes infinitely large. Thus, we can readily use the following asymptotic expansion of the Parabolic Cylinder function~\cite{hand1}: 
\begin{eqnarray}
D_{-n-1/2}(y)&\propto& y^{-n-1/2}\lbrack1 + 0(1/y)\rbrack \,\exp\left(-\frac{y^2}{4}\right), \nonumber \\
          y&\rightarrow & \infty,  \label{a16}
\end{eqnarray}
to rewrite the anharmonic one-phonon propagator~(\ref{a13}) as:
\begin{equation}
\mathcal{D}_{\lambda}(\omega_m)\rightarrow \mathcal{D}_0(\omega_m). \label{a17} 
\end{equation}
The well-known harmonic result for the phonon-mediated electron-electron interaction matrix then follows:
\begin{eqnarray}
g_0(\omega_m)&=& \alpha^2 \, {D}_0(\omega_m) \nonumber \\ 
&=&-\alpha^2/M(\omega_m^2 + \omega_0^2). \label{a18} 
\end{eqnarray}
In the classical regime i.e. $m=0$, $g_0(0)$ is negative implying a retarded attractive electron-electron interaction. \\
In the general case when the anharmonicity coefficient $\lambda$ is of arbitrary strength, the physics behind formula~(\ref{a15}) is not easy to capture from simple analytical considerations. Proceeding by numerical analysis, it is useful for a better understanding of the role played by anharmonic phonons in the two-electron correlations reflected by the interaction action~(\ref{a14})-(\ref{a15}), to first examine the salient features of the one-phonon propagator~(\ref{a13}). On figure~\ref{f1}, we plot $D_{\lambda}(\omega_m)/D_0(0)$ versus the reduced real frequency $i\omega_m/\omega_0$ for different values of the effective anharmonicity coefficient $\lambda_{eff}= \tilde{\lambda}/(M\omega_o^2)^2$. 
When $\lambda_{eff}=0$, the one-phonon propagator is singular at $i\omega_m=\omega_0$ in agreement with established behaviour of the free-phonon propagator. However, when the effective anharmonicity coefficient becomes nonzero the singularity is completely suppressed leading to optical phonon states of finite lifetimes~\cite{chow}. As we increase $\lambda_{eff}$, the finite tail of double peaks in the phonon propagator at the characteristic frequency $\omega_0$ decreases and vanishes at relatively strong anharmonicities relative to the characteristic harmonic-phonon stifness $M\omega_0^2$. Figure~(\ref{f2}) displays the resulting curves showing the variation of $g_{\lambda}(\omega_m, T)/g_0(0)$ as a function of $\lambda_{eff}$, for $i\omega_n=0$, $0.25\,\omega_0$, $0.5\,\omega_0$, $0.75\,\omega_0$ and $\omega_0$(from bottom to the top curves).   \\ 
Instructively, in addition to the pure harmonic regime consistently described for $\lambda=0$, formula~(\ref{a15}) also suggests two other distinct physical contexts involving a vanishing effective electron-electron interaction $g_{\lambda}(\omega_m, T)$. They are $L\rightarrow \infty$ but $T$ finite, and $T=0$ but $L$ finite. These two characteristic features, which are proper to the exact treatment of the anharmonic phonons, give evidence of the contribution of material sizes to electronic correlations mediated by anharmonic lattice vibrations.
\section{Anharmonic-phonon-induced CDW instability}
\begin{figure}[tb]
\begin{center}
\includegraphics{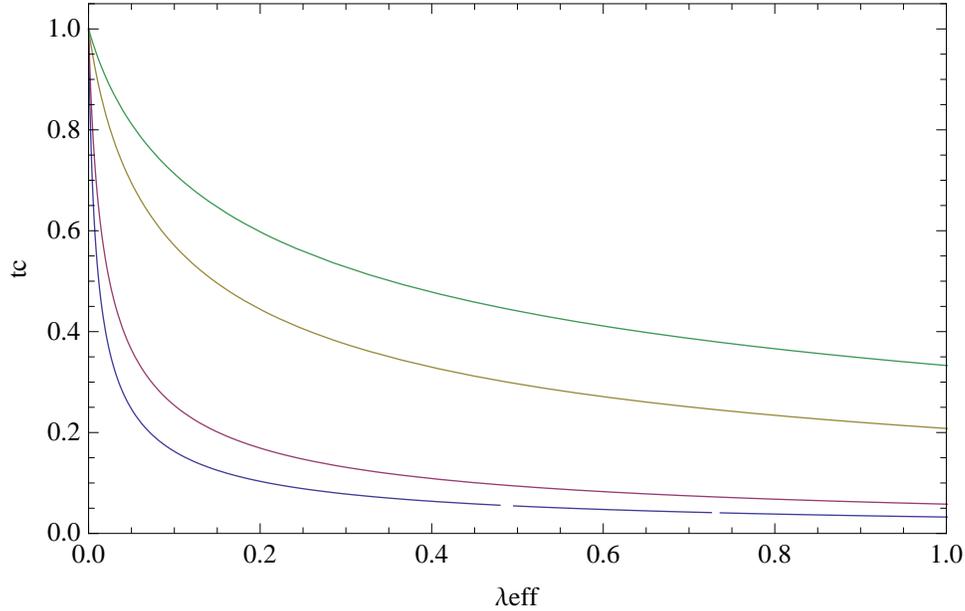}
\end{center}
\caption{The reduced CDW critical temperarture versus the effective anharmonicity coefficient $\lambda_{eff}$, for $\alpha^2/M\omega_0^2 =0.05$, 0.1, 0.5 and 1(from bottom to top curves).}
\label{f3}
\end{figure}

In this section we shall consider the relevant problem of CDW instability which has been widely studied within the framework of the Holstein model with harmonic lattice vibrations, and examine the effect of phonon anharmonicity on the transition temperature. Proceeding with, we follow the effective-field approach which involves constructing an effective Landau-Ginzburg(LG) functional associate with CDW fluctuations governed by the effective two-electron action~(\ref{a14}). First, we define the CDW operator: 
\begin{equation}
 \mathcal{O}(q, \tau)=\frac{1}{\sqrt{Na}}\sum_k{{\psi}^{\star}(k,\tau)\psi(k+q,\tau)} \label{a19}
\end{equation}
which helps rewrite the effective interaction action in a suitable form i.e.: 
\begin{equation}
S_I= -\frac{1}{2}\sum_{q}{\int_0^{\beta}{d\tau' \int_0^{\beta}{d\tau g_{\lambda}(\tau-\tau')\,\mathcal{O}(q, \tau)\mathcal{O}(-q, \tau')}}}.   \label{a20}
\end{equation}
Next, introducing appropriate Hubbard-Stratonovich fields conjugate to the above CDW oprators, the total partition function~(\ref{a2}) becomes:
\begin{eqnarray}
Z&=& Z_l^0\,\int{\prod_{k,\sigma,\tau} D\psi_{k,\sigma}(\tau)\,D\psi_{k,\sigma}^{\star}(\tau)\,D\phi(q,\tau)\,\,e^{S_{e}^0(\psi,\psi^{\star})}} \nonumber \\ 
&\times& \exp{\{ -\sum_q{\int_0^{\beta}{d\tau' \int_0^{\beta}{d\tau \phi(q, \tau) \lbrack 2g_{\lambda}^{-1}(\tau-\tau')\rbrack \phi(-q, \tau')}}}\}} \nonumber \\
&\times& \exp{\{-\sum_q {\int_0^{\beta}{d\tau \lbrack \phi(q, \tau)\mathcal{O}(-q, \tau) + \phi(-q, \tau)\mathcal{O}(q, \tau)\rbrack }}\}}. \label{a21}
\end{eqnarray}
Upon total elimination of electronic degrees of freedom and use of the linked cluster theorem, the last expression reduces to:
\begin{eqnarray}
Z&=&Z^0\, e^{-\beta\,\mathcal{F}\lbrack\phi\rbrack}, \nonumber \\
Z^0&=& Z_l^0\,Z_e^0, \label{a22}
\end{eqnarray}
where
\begin{equation}
\beta\,\mathcal{F}\lbrack\phi\rbrack= \sum_{q, \omega_m}{\lbrack 2g_{\lambda}^{-1}(\omega_m, T) - \chi(\omega_m, T)\rbrack}\, \vert \phi(q, \omega_m)\vert^2 + ... \label{a23}
\end{equation}
is the LG functional in which the CDW-CDW correlation function:
\begin{equation}
\chi(\omega_m, T)=  -\frac{2T}{L}\sum_{k, \omega_n}{G_0(k,\omega_n)G_0(k-2k_F,\omega_n - \omega_m)}, \hskip 0.4truecm \omega_n=(2n+1)\beta^{-1}, \label{a24}
\end{equation}
has been defined and is explicitely obtained for the nearly-free electron system as~\cite{dika3,jerome}
\begin{equation}
\chi(\omega_m, T) = N_F\left[\ln \frac{1.13 \, E_F}{T} + \it{\psi}\left(\frac{1}{2}\right)- \it{\psi} \left(\frac{1}{2} + \frac{\omega_m}{4 \pi T}\right)\right], \label{a25}
\end{equation}
with $N_F=1/2\pi \upsilon_F$, $\it{\psi}$ the Digamma function and $E_F=\upsilon_F k_F$ the Fermi energy. In the classical regime(i.e. $\omega_m=0$), the soft-mode condition requires the vanishing of coefficient of the Gaussian term in the LG functional~(\ref{a23}). To extract the critical temperature from this condition it is convenient to introduce $T_c^0$ for the harmonic-phonon Holstein model i.e.:
\begin{equation} 
T_c^0= 1.13\,E_F\,\exp{\frac{1}{\tilde{g_0}}}, \hskip 0.4truecm \tilde{g}_0= N_Fg_0(0). \label{a26}
\end{equation}   
With the help of this parameter we can define a reduced CDW transition temperature $t_c=T_c^{\lambda}/T_c^0$, and by numerical root finding we extract the $\lambda_{eff}-t_c$ phase diagram shown in figure~\ref{f3}. More precisely, on the figure $t_c$ is plotted versus $\lambda_{eff}$ for four representative values of the bare electron-phonon interaction coefficient $g_0(0)$, assumed to reflect distinct regimes of the electron-phonon interaction. 
A most remarkable fact emerging from curves is that $t_c$ is always a decreasing function of $\lambda_{eff}$, irrespective of the strength of the electron-phonon interaction. Equally remarkable is the reduction of the CDW transition temperature which is manifestly as strong as the electron-phonon interaction is weak. Quite strickingly no re-entrance phase appears for all the representative values of the electron-phonon interaction considered, contrasting with predictions of the self-consistent Gaussian phonon approximation~\cite{jarell1,jarell2}. 
\section{conclusions}
We obtained the expression of the phonon-mediated two-electron action for the molecular-crystal model with quartic anharmonic phonons, valid at any order of anharmonicity strength. To achieve our goal, we followed a non-perturbative approach based on non-Gaussian partial functional integrations of anharmonic phonon degrees of freedom. As a practical illustration of the advantage for integrating out exactly the anharmonic phonons we considered the CDW problem, and shown that the account of quartic anharmonic phonons in the Holstein model gives rise to a picture of the transition dominated by the interplay of material's local structure including the volume. However, for fixed volume it emerged that the suppression of the CDW transition was as strong as the electron-phonon interaction was weak. No re-entrant process was observed in the weak, medium and strong electron-phonon coupling regimes in contradiction with a recent prediction based on the self-consistent Gaussian approximation~\cite{jarell1,jarell2}. \\
From a general standpoint, anharmonic lattices display a wealth of rich properties that are distinct from their harmonic counterparts as established at length in past theoretical as well as experimental researches. Of fundamental interest, multi-phonon correlations lead to unusual phenomena such as biphonon boundstates, a broadening of one-phonon modes which is not a simple shift of mode frequencies in Raman spectra and the appearance of a distinctive structure of the phonon spectrum dominated by modes of finite lifetimes. Unveiling these properties requires a non perturbative treatment of anharmonic lattice vibrations in view of full understanding of their contributions to structural instabilities of anharmonic crystals. On the other hand, the importance of anharmonicity in electronic phase instabilities of low-dimensional materials has been demonstated in various experimental probes on lattice and electronic structures, including extended x-ray-absorption fine structures, neutron scattering and optical techniques. The lattice anharmonicity is thus observed as being particularly pronounced in planar structures such as cuprate oxydes and colossal manganites where it governs both intra and inter planar charge transfers. In most of these materials, the plane chain structure provides a suitable scenario for charge-transfer excitations owing to a possible reduction of the Coulomb repulsion due namely to interplane interactions, by anharmonic vibrations of ions along the chain(as in cuprate oxydes, see e.g.~\cite{bish2}). In this context the phonon anharmonicity can be so strong that the shift of the one-phonon frequency in the infrared Raman spectrum predicted within the self-consistent Gaussian renormalization approach, is not meaningful for true characterization of the temperature dependence of the effective charge-transfer interaction, the contribution of the local structure to structural instabilities and so on~\cite{bish2,bish3} as observed experimentally in real anharmonic materials. \\
Note to end that in recent years, there has been a great deal of interest~\cite{schulz,schulza,schulzb,voit} to the interplay of phonon-mediated attraction and Coulomb repulsion between electrons in Luttinger liquids. In particular, the Holstein-Hubbard model has been shown to display a phase diagram characterized by a great variety of groundstates including Mott insulator, charge-density-wave, spin-density-wave, s-wave and d-wave superconducting groundstates. For anharmonic materials belonging to this class, an exact formulation of the anharmonic phonon contribution to electronic correlations represents a relevant prerequisite towards a best account of specific features of their phase diagrams.  
\section*{Acknowledgment}
This work is part of a project done at the Max Planck Institute for the Physics of Complex Systems(MPIPKS), Dresden Germany within the visitor program award of the Institute(February-April 2009). The author is grateful to the Physical Society of Japan for financial support in publication.

\end{document}